\documentclass[a4paper,11pt]{article}
\pdfoutput=1 % if your are submitting a pdflatex (i.e. if you have
\usepackage{jheppub_BLNK} % for details on the use of the package, please
\usepackage[T1]{fontenc} % if needed
\usepackage{float}
\usepackage{epstopdf}
\usepackage{amsmath}

\title{\boldmath Lorentz symmetry violating BTZ black holes in massive gravity}

\author{Askar Ali and Khalid Saifullah} 
\affiliation{Department of Mathematics, Quaid-i-Azam University, Islamabad, Pakistan}

\emailAdd{askarali@math.qau.edu.pk} \emailAdd{ksaifullah@fas.harvard.edu}

\abstract{BTZ black holes provide excellent frameworks for studying theories that are at the interface of classical and quantum gravity. In this paper we couple the Riemannian spacetime with the bumblebee field, in the background of massive gravity, which produces the violation of spontaneous Lorentz symmetry. In this setup we construct a large family of static vacuum BTZ black hole solutions. We study the asymptotic behaviour of curvature invariants and show that our resulting solutions describe asymptotically AdS (2+1)-dimensional BTZ black holes with negative cosmological constant. For positive cosmological constant they are de Sitter. Thermodynamics of these black holes is also analysed.  
\vspace{90 mm}
}

\notoc 

\begin{document}
\maketitle
%\flushbottom 

%Key Words: Topological black holes; Hawking temperature; Scalar hair; Yang-Mills magnetic charge. 

\section{Introduction}
\label{sec:intro}
The general theory of relativity and the standard model of particle physics are two field theories that have described nature successfully. The former explains gravitation at classical level while the latter is a description of fundamental particles and interactions at the quantum level. In order to have a complete understanding of nature these two theories must be unified. Many quantum gravity theories have been developed in order to achieve this unification but these cannot be tested directly because of the present experimental limits on energy scale. 

There is another possibility that some signals of quantum gravity may be observed at very low energy scales of observations and their consequences could be examined in experimental works. One major signal among these is related to the violation of Lorentz symmetry. The idea of Lorentz symmetry breaking is very interesting and it has played an important role in string theory \cite{1,2}, loop quantum gravity theory \cite{4} and noncommutative field theories \cite{3}. The standard model extension is an effective field theory which studies gravity and the standard model at low energy scales. Apart from this,  additional terms are also contained in its structures which are responsible for information related to the Lorentz symmetry violation at the Planck scale \cite{5}. The electromagnetic sector of the standard model extension has been discussed in literature \cite{6,7}. There has been discussion on different facets of the electroweak sector \cite{8}, the strong sector \cite{9} and the hadronic physics \cite{10} also. The influence of Lorentz symmetry breaking on the gravitational sector has also been studied \cite{11,12,13,14}. For example, the case of gravitational waves was also explored in this context \cite{15}. Black holes have also been studied in the presence of Lorentz symmetry breaking. In this context, the static spherically symmetric vacuum black hole solution in the background of a bumblebee vector field has also been constructed \cite{16}. Thermodynamics of the black hole solution of such a bumblebee gravity was also studied \cite{17}. 

The effects and contributions of massive gravity have also been investigated in the context of astrophysics. For instance, the maximum mass of neutron stars greater than $3M_{sun}$ was found in a massive gravity model \cite{18}. In addition to this it is also seen that massive gravity can alter the thermodynamic properties of black holes \cite{19,20,21,22,23,24,25}. In particular, a behaviour similar to van der Waals' in the context of non-spherical black holes \cite{26}, Hawking temperature \cite{27} and their anti-evaporation was also investigated \cite{28,29}. Furthermore, massive gravitons in relation to  gravitational waves during inflation has also been discussed \cite{30}. The theory describing the possible free massive graviton was first analysed by Fierz and Pauli \cite{31,32}. Later, it was shown that there is a Boulware-Deser ghost instability in this massive gravity theory at nonlinear level \cite{33,34}. However, lately a significant approach was used to construct a massive gravity theory where this type of ghost instability was absent \cite{35}. A simple method to build a massive gravity theory is to add the mass term in the Einstein-Hilbert Lagrangian. If we assume that the mass of graviton is $m$ then Einstein's theory is recovered in the limit $m\rightarrow 0$. Here, in this work, we use a massive gravity theory that was introduced by Vegh \cite{36}. 

General relativity in (2+1)-dimensions has become a very famous model at both classical and quantum levels \cite{37}. However, one of the major drawbacks of these (2+1)-dimensional models is that there is no Newtonian limit \cite{38}. In 1992, Ba\~{n}ados, Teitelboim and Zanelli (BTZ) presented a very interesting black hole solution \cite{39} in (2+1)-dimensional gravity with the negative cosmological constant. The AdS charged black hole solution of Einstein-Maxwell theory corresponds to the charged BTZ black hole \cite{40,41}. In this work, we construct a large class of vacuum static circularly symmetric (2+1)-dimensional black holes in massive bumblebee gravity. In this context, we construct solutions which depend on the bumblebee parameter, and for each real value of this parameter we will get a new solution of massive gravity in the presence of Lorentz symmetry breaking. Having constructed these BTZ solutions, we also investigate their thermodynamics, namely, we compute several thermodynamic quantities and study their behaviour. The results obtained in the present paper are different from the usual Einstein's general relativity and massive gravity theories as well. 

The paper is planned in the following fashion. In Section 2, the massive gravity is coupled to the bumblebee vector field and the corresponding modified gravitational field equations are derived. The exact static circularly symmetric vacuum BTZ black hole solutions of this gravity theory are calculated. We also discuss geometry of the black hole and asymptotic behaviour of curvature invariants. In Section 3, thermodynamics of the resulting black hole solution is studied and several relevant quantities computed. Finally, we conclude our work in Section 4.

\section{(2+1)-dimensional black holes of massive bumblebee gravity} 

The action function associated with (2+1)-dimensional massive gravity in the presence of bumblebee vector field $B_{\mu}$ can be written in the form \cite{17,19}
\begin{eqnarray}\begin{split}
I&=\int d^3x\sqrt{-g}\bigg[\frac{R}{2\kappa}-2\Lambda+\frac{\xi}{2\kappa}B^{\mu}B^{\nu}R_{\mu\nu}-\frac{1}{4}B_{\mu\nu}B^{\mu\nu}-V(B^{\mu})+L_M\\&+m^2\Sigma_{i=1}^3c_iU_i(g,f)\bigg],
\label{1}\end{split}
\end{eqnarray}
where $R$ is the Ricci scalar, $\Lambda$ is the cosmological constant, $g$ is the metric tensor and $f$ is the fixed symmetric tensor. Furthermore, $\xi$ is the real constant which describes the nonminimal coupling of gravity with the bumblebee field and $L_M$ is the Lagrangian density corresponding to the matter contents. The $B_{\mu\nu}$ represents the strength of the bumblebee vector field and is given by
\begin{equation}
B_{\mu\nu}=\partial_{\mu}B_{\nu}-\partial_{\nu}B_{\mu}.\label{2}
\end{equation}
Also, in (\ref{1}) $c_is'$ are constants and $U_is'$ are symmetric polynomials of the eigenvalues of the $3\times 3$ matrix, $K^{\mu}_{\nu}=\sqrt{g^{\mu\alpha}f_{\alpha\nu}}$, which are written as
\begin{eqnarray}\begin{split}
U_1=[K], U_2=[K]^2-[K^2], 
U_3=[K]^3-3[K][K^2]+2[K^3].\label{3}\end{split}
\end{eqnarray}
Let's choose a potential $V$ in the general functional form provided that it gives a nonvanishing vacuum expectation value for the bumblebee field $B_{\mu}$
\begin{equation}
V=V(B_{\mu}B^{\mu}\pm b^2),\label{4}
\end{equation}
where $b$ is a positive real constant. The expectation value is obtained when $V(B_{\mu}B^{\mu}\pm b^2)=0$, so that the constraint 
\begin{equation}
B_{\mu}B^{\mu}=\mp b^2,\label{5}
\end{equation}
is satisfied. This is easily solvable when the non-null expectation value of a bumblebee field is written in the form
\begin{equation}
E(B_{\mu})=b_{\mu},\label{6}
\end{equation}
where the vector $b_{\mu}$ depends on the spacetime coordinates in such a way that $b_{\mu}b^{\mu}=\mp b^2=$constant. Due to this non-null background of $b_{\mu}$, the Lorentz symmetry is completely violated. It should be noted that the negative and positive signs in (\ref{5}) indicate the timelike or spacelike behaviour of the vector field, respectively.
 
 Using the variational principle we find that the variation relative to the metric tensor gives the following modified Einstein field equations from action (\ref{1})
\begin{eqnarray}
\label{7}
\begin{split}
R_{\mu\nu}-\frac{1}{2}Rg_{\mu\nu}+\Lambda g_{\mu\nu}+m^2X_{\mu\nu}&=\kappa T^{(M)}_{\mu\nu}+\kappa T^{(B)}_{\mu\nu},
\end{split}
\end{eqnarray}
where $g_{\mu\nu}$ is the metric tensor, $R_{\mu\nu}$ is the Ricci tensor, $T^{(B)}_{\mu\nu}$ is the energy-momentum tensor corresponding to the bumblebee vector field defined as
\begin{eqnarray}
\label{8}
\begin{split}
T^{(B)}_{\mu\nu}&=-B_{\mu\alpha}B^{\alpha}_{\nu}-\frac{1}{4}B_{\alpha\beta}B^{\alpha\beta}g_{\mu\nu}-Vg_{\mu\nu}+2V'B_{\mu}B_{\nu}\\&+\frac{\xi}{\kappa}\bigg[\frac{1}{2}B^{\alpha}B^{\beta}R_{\alpha\beta} g_{\mu\nu}-B_{\mu}B^{\alpha}R_{\alpha\nu}+\frac{1}{2}\partial_{\alpha}\partial_{\mu}\bigg(B^{\alpha}B_{\nu}\bigg)\\&+\frac{1}{2}\partial_{\alpha}\partial_{\nu}\bigg(B^{\alpha}B_{\mu}\bigg)-\frac{1}{2}\partial^{\alpha}\partial_{\alpha}\bigg(B_{\mu}B_{\nu}\bigg)-\frac{1}{2}g_{\mu\nu}\partial_{\alpha}\partial_{\beta}\bigg(B^{\alpha}B^{\beta}\bigg)\bigg],
\end{split}
\end{eqnarray}
and
\begin{eqnarray}
\label{9}
\begin{split}
X_{\mu\nu}&=-\frac{c_1}{2}\bigg(U_1g_{\mu\nu}-K_{\mu\nu}\bigg)-\frac{c_2}{2}\bigg(U_2g_{\mu\nu}-2U_1K_{\mu\nu}+2K^2_{\mu\nu}\bigg)-\frac{c_3}{2}\\&\times\bigg(U_3g_{\mu\nu}-3U_2K_{\mu\nu}+6U_1K^2_{\mu\nu}-6K^3_{\mu\nu}\bigg).
\end{split}
\end{eqnarray}
Notice that primes denote derivatives with respect to the argument. In the same manner, by varying with respect to the bumblebee field, Eq. (\ref{1}) gives the following equation of motion corresponding to the bumblebee field 
\begin{equation}
\partial^{\mu}B_{\mu\nu}=J_{\nu},\label{10}
\end{equation}
where $J_{\nu}=J^{(B)}_{\nu}+J^{(M)}_{\nu}$. Here, $J^{(M)}_{\nu}$ is related to the matter sector and $J^{(B)}_{\nu}$ comes into play due to the self interaction of the bumblebee field and is defined as
\begin{equation}
J^{(B)}_{\nu}=2V'B_{\nu}-\frac{\xi}{\kappa}B^{\mu}R_{\mu\nu}.\label{11}
\end{equation}
As we want to calculate the static circularly symmetric black hole solutions of (\ref{7}), therefore, we assume a line element of the form
\begin{equation}
ds^{2}=-f(r)dt^2+\frac{dr^2}{f(r)}+r^2d\theta^2,\label{12}
\end{equation}
where $f(r)$ is the metric function which is to be determined. It is worth mentioning that in this paper we focus on a vacuum solution, therefore, we are choosing the matter tensor $T^{(M)}_{\mu\nu}$ to be zero. It is clearly seen that the bumblebee potential in (\ref{1}) vanishes when (\ref{6}) is satisfied, a property that characterizes the vacuum. In other words, we are interested in vacuum in the presence of Lorentz symmetry breaking, so the bumblebee field remains stationary and is determined by its expectation value $b_{\mu}$. Similar hypothesis was also considered in \cite{16,17} for the characterization of vacuum under the influence of the bumblebee vector field. Therefore, the bumblebee field is fixed to be
\begin{equation}
B_{\mu}=b_{\mu}, \label{13}
\end{equation}
and as a consequence, we have $V=0$ and $V'=0$. Now, consider a spacelike background $b_{\mu}$ by choosing the following form
 \begin{equation}
 b_{\mu}=\bigg(0,b_r(r),0\bigg).\label{14}
 \end{equation}
 Further, after we have considered the background field of the form (\ref{14}), then as a consequence all the components $b_{\mu\nu}$ would vanish. With the help of the condition $b_{\mu}b^{\mu}=\mp b^2=$constant, we can find out the following expression for the background radial field
 \begin{equation}
 b_r(r)=\frac{|b|}{\sqrt{f(r)}}.\label{15}
 \end{equation} 
  
\begin{figure}[h]
	\centering
	\includegraphics[width=0.8\textwidth]{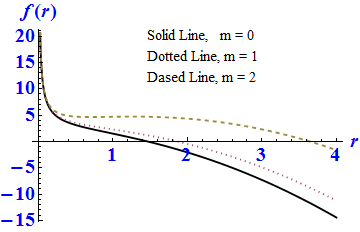}
	\caption{Plot of function $f(r)$ (Eq. (\ref{16})) vs $r$ for fixed values of $\mu=-1.0$, $l=0.5$, $\beta=1.0$, $k=1$, $c_1=1.0$ and $\Lambda=-2.0$.}\label{Askar1}
\end{figure}

  We choose the metric of the form $f_{\mu\nu}=diag(0,0,\beta^2)$ where $\beta$ is a positive constant. This choice of reference metric will enable us to determine the polynomials $U_i$ in the form $U_j=\frac{C^j}{r^j}\prod_{x=2}^{j+1}(d-x)$. Thus, by using the above information and line element (\ref{12}) in the field equations (\ref{7}), the required metric function is evaluated as
\begin{eqnarray}
\begin{split}
\label{16} 
f(r)&=k-\frac{3\mu l}{2(l-1)r^{\frac{2(1-l)}{3l}}}-\frac{\Lambda r^2}{2l+1}-\frac{2m^2\beta c_1r}{l+2},
\end{split} 
\end{eqnarray}
where $\mu$ is the integration constant that is related to ADM mass of the (2+1)-dimensional black hole. Here $l$ is the real parameter related to the bumblebee field such that $l=\xi b^2$. We can find the event horizons of the black hole by solving the equation $f(r)=0$. Fig. (\ref{Askar1}) shows that the value of the radial variable at which the graph intersects the horizontal axis points out the position of the event horizon. 
 
Now, the Ricci scalar and the Kretschmann scalar for the metric (\ref{12}) are given by
 
 \begin{eqnarray}\begin{split}
 R(r)&=-\frac{2}{r}\bigg(\frac{df}{dr}\bigg)-\frac{d^2f}{dr^2},\label{17}\end{split}
 \end{eqnarray}
 and
  \begin{eqnarray}\begin{split}
 K(r)&=\frac{2}{r^2}\bigg(\frac{df}{dr}\bigg)^2+\frac{d^2f}{dr^2}.\label{18}\end{split}
 \end{eqnarray}
  Differentiation of Eq. (\ref{16}) yields 
   \begin{eqnarray}\begin{split}
  \frac{df}{dr}&=\frac{2m^2\beta c_1}{l+2}-\frac{4\Lambda r}{4l+2}+\frac{\mu}{r^{\frac{l+2}{3l}}}.\label{19}\end{split}
  \end{eqnarray}
  Differentiating it again, we get
  \begin{eqnarray}\begin{split}
  \frac{d^2f}{dr^2}&=-\frac{4\Lambda }{4l+2}-\frac{\mu(l+2)}{3lr^{\frac{4l+2}{3l}}}.\label{20}\end{split}
  \end{eqnarray}
  Therefore, by using Eqs. (\ref{16})-(\ref{20}), we obtain
  \begin{equation}
  R(r)=-\frac{4m^2\beta c_1}{(l+2)r}+\frac{12\Lambda}{4l+2}+\frac{\mu(2-5l)}{3lr^{\frac{4l+2}{3l}}}, \label{21}
  \end{equation}
  and
  \begin{eqnarray}\begin{split}
   K(r)&=\frac{8m^4c_1^2\beta^2}{(l+2)^2r^2}+\frac{48\Lambda^2}{(4l+2)^2}+\frac{2\mu^2}{r^{\frac{8l+4}{3l}}}+\frac{8m^2\beta c_1\mu}{(l+2)r^{\frac{2-5l}{3l}}}\\&-\frac{16\Lambda\mu}{(4l+2)r^{\frac{4l+2}{3l}}}-\frac{32\Lambda m^2c_1\beta}{(l+2)(4l+2)r}+\frac{\mu^2(l+2)^2}{9l^2r^{\frac{8l+4}{3l}}}\\&+\frac{8\Lambda\mu (l+2)}{3l(4l+2)r^{\frac{4l+2}{3l}}}. \label{22}\end{split}
  \end{eqnarray}
  It is clearly seen that for $2l+1>0$, the Ricci scalar becomes infinite as $r\rightarrow 0$ and $R(r)=12\Lambda/(4l+2)$ as $r\rightarrow\infty$. Also, the Kretschmann scalar is undefined as $r\rightarrow 0$ and attains the value $K(r)=48\Lambda^2/(4l+2)^2$ as $r\rightarrow\infty$. Thus, the asymptotic behaviour of Ricci and Kretschmann scalars indicates that our solution (\ref{16}) describes the AdS BTZ black hole solution of massive bumblebee gravity with negative $\Lambda$ in the presence of spontaneous Lorentz symmetry breaking. This will be de Sitter when $\Lambda$ is positive. It should be noted that for $\Lambda=0$, (\ref{16}) yields the asymptotically flat black hole solutions of massive bumblebee gravity. Furthermore, for $m=0$, the metric function (\ref{16}) reduces to 
  \begin{eqnarray}
  \begin{split}
  \label{22a} 
  f(r)&=k-\frac{3\mu l}{2(l-1)r^{\frac{2(1-l)}{3l}}}-\frac{\Lambda r^2}{2l+1},
  \end{split} 
  \end{eqnarray}
  which describes the vacuum static (2+1)-dimensional black hole solution in this gravity theory. One can easily verify it from action (\ref{1}) and field equations (\ref{7}) under the same process for $m=0$. In other words, this is a BTZ black hole solution of gravitational field equations determined from coupling of the bumblebee vector field with Einstein's theory of gravity. The corresponding four-dimensional static and spherically symmetric black hole solution is studied in \cite{16,17}.
  
\section{Thermodynamics of (2+1) black holes} 

Here, in this section we want to compute some of the thermodynamic quantities of the bumblebee black hole solutions obtained in the previous section and check the validity of the first law of thermodynamics. Note that we are using the definition of Hawking temperature $T_H$ which is obtained from the notion of surface gravity at the outer horizon $r_+$. Hence, we find 
\begin{eqnarray}
\label{23}
\begin{split} 
T_H&=\frac{m^2\beta c_1}{2\pi (l+2)}-\frac{\Lambda r_+}{\pi (4l+2)}+\frac{\mu}{4\pi r_+^{\frac{l+2}{3l}}}. 
\end{split} 
\end{eqnarray}
The ADM mass of the black hole can easily be obtained in terms of the outer horizon from equation $f(r_+)=0$. Thus we have
 \begin{eqnarray}
 \label{24}
 \begin{split} 
 \mu&=\frac{2\Lambda (l-1)r_+^{\frac{4l+2}{3l}}}{3l(2l+1)}-\frac{2(l-1)kr_+^{\frac{2-2l}{3l}}}{3l}-\frac{4m^2\beta c_1(l-1)r_+^{\frac{l+2}{3l}}}{3l(l+2)}. 
 \end{split} 
 \end{eqnarray}
Hence, with the help of this value of finite mass in terms of horizon radius, the Hawking temperature (\ref{23}) takes the following form
\begin{eqnarray}
\begin{split}
T_H&=\frac{m^2\beta c_1}{6\pi l}-\frac{\Lambda r_+}{6\pi l}-\frac{k(l-1)}{6\pi l r_+}.\label{25}\end{split}
\end{eqnarray}
\begin{figure}[h]
	\centering
	\includegraphics[width=0.8\textwidth]{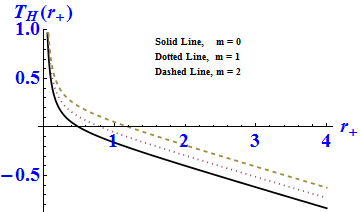}
	\caption{Plot of function $T_H(r_+)$ (Eq. (\ref{25})) vs $r_+$ for fixed values of $l=0.5$, $\beta=1.0$, $k=1$, $c_1=1.0$ and $\Lambda=-2.0$.}\label{Askar2}
\end{figure}
From the above temperature, the black hole entropy can be obtained by making use of the thermodynamic relation $T_HdS=d\mu$ which yields
 \begin{eqnarray}
 \begin{split}
 S&=\int\frac{d\mu}{T_H(r_+)}=\frac{2}{3l}\int\frac{\bigg(4\Lambda(l-1)r_+^{\frac{l+2}{3l}}+4(l-1)^2kr_+^{\frac{2-5l}{3l}}-4m^2\beta c_1(l-1)r_+^{\frac{2-2l}{3l}}\bigg)\pi r_+}{m^2c_1\beta r_+-k(l-1)-\Lambda r_+^2}.\label{26}\end{split}
 \end{eqnarray}
If we use these thermodynamic quantities, we can easily see from Eqs. (\ref{23})-(\ref{26}) that the first law of black hole thermodynamics is satisfied by virtue of the relation $T_H=d\mu/dS$.
Now the heat capacity is given by
\begin{equation}
C=T_H\frac{\partial S}{\partial T_H}=T_H\frac{\partial S/\partial r_+}{\partial T_H/\partial r_+}.\label{27}
\end{equation}
Thus, using value of the Hawking temperature (\ref{25}) and the above equation (\ref{26}), we find the expression of heat capacity to be given by 
\begin{eqnarray}
\begin{split}
C&=\frac{16\pi(l-1)}{k(l-1)-\Lambda r^2}\bigg[\Lambda r_+^{\frac{7l+2}{3l}}+(l-1)kr_+^{\frac{2+l}{3l}-4m^2c_1\beta r_+^{\frac{2+4l}{3l}}} \bigg].\label{28}
\end{split}
\end{eqnarray}
\begin{figure}[h]
	\centering
	\includegraphics[width=0.8\textwidth]{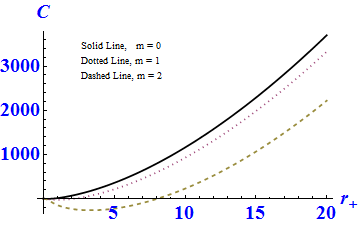}
	\caption{Plot of function $C_H(r_+)$ (Eq. (\ref{28})) vs $r_+$ for fixed values of $l=0.5$, $\beta=1.0$, $k=1$, $c_1=1.0$ and $\Lambda=-2.0$.}\label{Askar3}
\end{figure}
Note that the heat capacity gives information which is useful in the analysis of thermodynamic stability and instability of black holes. The black hole is unstable in those regions where this quantity changes sign from positive to negative. For example, the curve of heat capacity for $m=2$, in Fig. (\ref{Askar3}) shows that at $r_+=8.0$ heat capacity changes sign, so this corresponds to the first order phase transition and in the interval $0<r_+<8$ the black hole is unstable. Furthermore, the point where this quantity reaches infinite value, gives second order phase transition of the black hole. The second order phase transition corresponds to the value $r_+=\sqrt{k(l-1)/\Lambda}$, since at this value $\partial T_H/\partial r_+$ vanishes. 

Now, let us compute the free energy of the system which tells us about the amount of work done by a thermodynamic system. The unusable energy which cannot perform any work is the combination of temperature and total entropy of the system. Therefore, for our black hole solution we have the following Helmholtz free energy
   \begin{eqnarray}
  \begin{split}
  F&=\mu-T_HS=\frac{2\Lambda(l-1)r_+^{\frac{4l+2}{3l}}}{3l(2l+1)}-\frac{2(l-1)kr_+^{\frac{2-2l}{3l}}}{3l}-\frac{4m^2\beta c_1(l-1)r_+^{\frac{l+2}{3l}}}{3l(l+2)}+\frac{4(l-1)}{9l^2}\\&\times\bigg(m^2\beta c_1-\Lambda r_+-\frac{k(l-1)}{r_+}\bigg)\int_{0}^{r_+}\frac{\Lambda x^{\frac{4l+2}{3l}}+k(l-1)x^{\frac{2-2l}{3l}}-4m^2\beta c_1x^{\frac{l+2}{3l}}}{\Lambda x^2-m^2\beta c_1x+k(l-1)}dx.\label{29}
  \end{split}
  \end{eqnarray}

\section{Conclusion} 

In this paper, we have studied the (2+1)-dimensional spacetime geometry in the presence of a bumblebee vector field. In this context, we have investigated a static vacuum circularly symmetric black hole solution of massive gravity with spontaneous Lorentz symmetry breaking. This symmetry is violated due to the coupling of bumblebee vector field with the massive gravity. It is shown that the obtained solution reduces to the vacuum solution (\ref{22a}) of bumblebee gravity for the vanishing massive parameter i.e. when $m=0$. It should also be mentioned that when the cosmological constant $\Lambda$ is zero, our solution describes an asymptotically flat black hole. The asymptotic behaviour at both radial infinity and origin are also studied which show that the obtained (2+1)-dimensional spacetime geometry is non-asymptotically flat and that the gravitating object describing our solution possesses the central essential singularity at the origin.

The thermodynamics of the black hole is also studied and we work out different thermodynamic quantities such as Hawking temperature, entropy,  and heat capacity for the black hole solution. From the corresponding expressions of Hawking temperature and heat capacity it is clear that for such a black hole, phase transitions occur. From Fig. (\ref{Askar2}), it is seen that the region where the curve corresponding to temperature gives negative values corresponds to the region of instability and the point at which the temperature curve intersects the horizontal axis corresponds to first order phase transition of the black hole. Furthermore, the points at which the heat capacity diverges indicates the second order phase transition. Heat capacity diverges at the critical points of temperature i.e. at which $\partial T_H/\partial r_+=0$. Thus, from (\ref{25}) and (\ref{28}), the second order phase transition takes place at $r_+=\sqrt{k(l-1)/\Lambda}$. 

\section*{Acknowledgements}
A research grant from the Higher Education Commission of Pakistan under its Project No. 6151 is gratefully acknowledged.


\begin{thebibliography}{99}
\bibitem{1} V. A. Kostelecky and S. Samuel, Phys. Rev. D \textbf{39} (1989) 683.
\bibitem{2} D. Colladay and V. A. Kostelecky, Phys. Rev. D \textbf{55} (1997) 6760. 
\bibitem{4} R. Gambini and J. Pullin, Phys. Rev. D \textbf{59} (1999) 124021. 
\bibitem{3} S. M. Carroll, J. Harvey, V. A. Kostelecky, C. D. Lane and T. Okamoto, Phys. Rev. Lett. \textbf{87} (2001) 141601. 
\bibitem{5} V. A. Kostelecky, Phys. Rev. D \textbf{69} (2004) 105009.
\bibitem{6} V. A. Kostelecky and R. Lehnert, Phys. Rev. D \textbf{63} (2001) 065008.
\bibitem{7} K. Bakke and H. Belich, Eur. Phys. J. Plus \textbf{129} (2014) 147.
\bibitem{8} D. Colladay and P. Mcdonald, Phys. Rev. D \textbf{79} (2009) 125019.
\bibitem{9} V. M. Abazov et al. (The D0 Collaboration), Phys. Rev. Lett. \textbf{108} (2012) 261603.
\bibitem{10} M. S. Berger, V. A. Kostelecky and Z. Liu, Phys. Rev. D \textbf{93} (2005) 036005.
\bibitem{11} R. Bluhm and V. A. Kostelecky, Phys. Rev. D \textbf{71} (2005) 065008.
\bibitem{12} A. F. Santos and S. C. Ulhoa, Mod.Phys.Lett. A \textbf{30} (2015) 1550011.
\bibitem{13} R. V. Maluf, V. Santos, W. T. Cruz and C. A. S. Almeida, Phys. Rev. D \textbf{88} (2013) 025005.
\bibitem{14} R. V. Maluf, C. A. S Almeida, R. Casana and M. Ferreira, Phys. Rev. D \textbf{90} (2014) 025007.
\bibitem{15} V. A. Kostelecky  and M. Mewes, Phys. Lett. B \textbf{757} (2016) 510.
\bibitem{16} R. Casana, A. Cavalcante F. P. Poulis and E. B. Santos, Phys. Rev. D \textbf{97} (2018) 104001. 
\bibitem{17} D. A. Gomes, R. V. Maluf and C. A. S. Almeida, Thermodynamics of Schwarzschild-like black holes in bumblebee gravity models, arXiv: 1811.08503 [gr-qc] (2018).
\bibitem{18} S. H. Hendi, G. H. Bordbar, B. E. Panah and S. Panahiyan, JCAP \textbf{07} (2017) 4.
\bibitem{19} A. Bouchareb and G. Clement, Class. Quantum Grav. \textbf{24} (2007) 5581.
\bibitem{20} F. Capela and P. G. Tinyakov, JHEP \textbf{04} (2011) 42.
\bibitem{21} M. S. Volkov, Lect. Notes Phys. \textbf{892} (2015) 161. 
\bibitem{22} E. Babichev and R. Brito, Class. Quantum Grav. \textbf{32} (2015) 154001.
\bibitem{23} S. G. Gosh, L. Tannukij and P. Wongjun, Eur. Phys. J. C \textbf{76} (2016) 119 .
\bibitem{24} M. Zhang and W. B. Liu, Coexistent physics of massive black holes in the phase transitions, arXiv:1610.03648 (2016).
\bibitem{25} Y. P. Hu, X. X. Zeng and H. Q. Zhang, Phys. Lett. B \textbf{765} (2017) 120.
\bibitem{26} S. H. Hendi, R. B. Mann, S. Panahiyan and B. E. Panah, Phys. Rev. D \textbf{95} (2017) 021501.
\bibitem{27} S. H. Hendi, B. E. Panah and S. Panahiyan, JHEP \textbf{05} (2016) 29.
\bibitem{28} T. Katsuragawa, Universe \textbf{1} (2015) 158.
\bibitem{29} T. Katsuragawa and S. Nojiri, Phys. Rev. D \textbf{91} (2015) 084001.
\bibitem{30} A. E. Gumrukcuoglu, S. Kuroyanagi, C. Lin, S. Mukohyama and N. Tanahashi, Class. Quantum Grav. \textbf{29} (2012) 235026.
\bibitem{31} M. Fierz, Helv. Phys. Acta \textbf{12} (1939) 3.
\bibitem{32} M. Fierz and W. Pauli, Proc. R. Soc. A \textbf{173} (1939) 211.
\bibitem{33} D. G. Boulware and S. Deser, Phys. Rev. D \textbf{6} (1972) 3368.
\bibitem{34} D. G. Boulware and S. Deser, Phys. Lett. B \textbf{40} (1972) 227.
\bibitem{35} C. de Rham and G. Gabadadze, Phys. Rev. D \textbf{82} (2010) 044020.
\bibitem{36} D. Vegh, Holography without translational symmetry, arXiv:1301.0537 (2013).
\bibitem{37} S. Carlip, J. Korean Phys. Soc. \textbf{28} (1995) S447.
\bibitem{38} J. D. Barrow, A. B. Burd and D. Lancaster, Class. Quantum. Grav. \textbf{3} (1986) 551.
\bibitem{39} M. Ba\~{n}ados, C. Teitelboim and J. Zanelli, Phys. Rev. Lett. \textbf{69} (1992) 1849.
\bibitem{40} S. Carlip, Class. Quantum. Grav. \textbf{12} (1995) 2853.
\bibitem{41} C. Martinez,  C. Teitelboim and J. Zanelli, Phys. Rev. D \textbf{61} (2000) 104013.

\end{thebibliography}
\end{document}